\documentclass[11pt]{article}

\usepackage{amsmath,amssymb,amsxtra}
\usepackage{natbib,verbatim}

\newtheorem{definition}{Definition}

\newcommand{\xvec}{{\bf x}}
\newcommand{\yvec}{{\bf y}}

\newcommand{\styp}{A_{\epsilon}^{*(n)}}

\newcommand{\mhY}{\mathcal{\hat{Y}}}
\newcommand{\mX}{\mathcal{X}}
\newcommand{\mY}{\mathcal{Y}}
\newcommand{\mS}{\mathcal{S}}

\newcommand{\mW}{\mathcal{W}}
\newcommand{\mL}{\mathcal{L}}

\newcommand{\hY}{\hat{Y}}
\newcommand{\hhY}{\hat{\hat{Y}}}

\newcommand{\hyvec}{\hat{{\bf y}}}
\newcommand{\hy}{\hat{y}}

\newcommand{\hw}{\hat{w}}

\newcommand{\Pe}{P_{e}^{(n)}}

\title{The Discrete Memoryless Relay Channel: Joint-Decoding vs.
  Sequential Decoding of~\cite[theorem 6]{CoverG:79}}
\author{Ron Dabora \hspace{2cm} Sergio D.\ Servetto}
\date{School of Electrical and Computer Engineering -- Cornell University}

\begin{document}
\maketitle

\vspace{-0.0cm}

\section{Definitions}

\begin{definition}
    \label{def:relay_channel}
    {\em
    The {\em discrete relay channel} is defined by two discrete input alphabets $\mX_1$ and $\mX_2$, two
    discrete output alphabets $\mY_1$ and $\mY$ and a probability density function $p(y,y_1|x_1,x_2)$ giving the
    probability distribution on $\mY \times \mY_1$ for each $(x_1,x_2) \in \mX_1 \times \mX_2$.
    The relay channel is called {\em memoryless} if the probability of a block of $n$ transmissions is given by
    $p(\yvec,\yvec_1|\xvec_1,\xvec_2) = \prod_{i=1}^n p\left(y_i, y_{1,i}|x_{1,i},x_{2,i}\right)$.}
\end{definition}
\begin{definition}
    \label{def:code}
    {\em
    A {\em $(2^{nR},n)$ code} for the relay channel consists of a source message set
    $\mW = \left\{1,2,...,2^{nR}\right\}$, a mapping function $f$ at the encoder,
    \[
        f: \mW \mapsto \mX_1^n,
    \]
    a set of $n$ relay functions
    \[
        x_{2,i} = t_i\left(y_{1,1},y_{1,2},...,y_{1,i-1} \right),
    \]
    where the $i$'th relay function $t_i$ maps the first $i-1$ channel inputs at the relay into a transmitted
    relay symbol at time $i$. Lastly we have a decoder
    \[
        g: \mY^n \mapsto \mW.
    \]}
\end{definition}
\begin{definition}
    \label{def:Perr}
    {\em
    The {\em average probability of error} of a code with length $n$ for the relay channel is defined as
    \[
        \Pe = \Pr(g(Y^n) \ne W),
    \]
    where $W$ is selected uniformly over $\mW$.}
\end{definition}

\section{Joint-Decoding at the Destination Receiver}
    The code construction is essentially the same as the one devised in \cite[theorem 6]{CoverG:79}
    and the same procedure for decoding and encoding at the relay is used. The main change is in the decoding at the receiver:

\subsection{Decoding at the Destination at Time $i$}
    \label{subsec:Dec_at_dest}
    At time $i$ the receiver decodes $w_{i-1}$.
    \begin{enumerate}
        \item From $\yvec(i)$, the received signal at time $i$, the receiver decodes $s_i$ by looking for a unique
        $s \in \mS$, the set of partition indices used to select $\xvec_2$, such that $\big(\xvec_2(s), \yvec(i)\big) \in \styp$.
        From the single channel capacity theorem, see \cite[Ch. 8.4]{Yeung:01},
        the correct $s_i$ can be decoded with an arbitrarily small probability of error by taking $n$ large
        enough as long as
        \begin{equation}
        \label{eqn:R0_conds}
            R_0 \le I(X_2;Y).
        \end{equation}

        \item The receiver now knows the set $S_{s_i}$ into which $m_{i-1}$ (the relay message at time $i-1$) belongs.
        Additionally, from decoding at time $i-1$
        the receiver knows $s_{i-1}$, used to generate $m_{i-1}$.

        \item The receiver generates the set
        $\mL(i-1) = \left\{ w \in \mW: \big(\xvec_1(w), \yvec(i-1), \xvec_2(s_{i-1})\big) \in \styp\right\}$.

        \item The receiver now looks for a unique $w \in \mL(i-1)$ such that
            $\big( \xvec_1(w), \yvec(i-1), \hyvec_1(m|s_{i-1}), \xvec_2(s_{i-1})\big) \in \styp$ for some
            $m \in S_{s_i}$.
            If such a unique $w$ exists then it is the decoded $\hw_{i-1}$,
            otherwise the receiver declares an error.
    \end{enumerate}
\section{Comparison of the Rate Obtained with Joint Decoding vs. the Sequential Decoding of
    \cite[theorem 6]{CoverG:79}}

    We now explain why joint-decoding does not improve on the EAF
   rate.\\
   The standard EAF expression is given by \cite[theorem 6]{CoverG:79}:
   \begin{eqnarray*}
        R & \le & I(X_1;Y,\hY_1|X_2) = I(X_1;Y|X_2) + I(X_1;\hY_1|X_2,Y)\\
        \mbox{subject to }I(X_2;Y) & \ge & I(\hY_1;Y_1|X_2,Y),
   \end{eqnarray*}
    where $p(x_1,x_2,y,y_1,\hy_1) = p(x_1)p(x_2)p(y,y_1|x_1,x_2)p(\hy_1|x_2,y_1)$.
   Joint decoding results in the following rate expression
   \begin{eqnarray*}
        R & \le & I(X_1;Y|X_2)  + \min(I(X_2;Y) - I(\hY_1;Y_1|X_1,X_2,Y),  I(X_1;\hY_1|X_2,Y))\\
        \mbox{subject to }I(X_2;Y) & \ge & I(\hY_1;Y_1|X_1,X_2,Y) = I(\hY_1;Y_1|X_2,Y) - I(X_1;\hY_1|X_2,Y),
   \end{eqnarray*}
   with $p(x_1,x_2,y,y_1,\hy_1) = p(x_1)p(x_2)p(y,y_1|x_1,x_2)p(\hy_1|x_2,y_1)$.

   Now fix $p(\hy_1|x_2,y_1)$, $p(x_1)$ and $p(x_2)$,
    and assume that $I(X_2;Y) \ge  I(\hY_1;Y_1|X_2,Y)$. Then

    \[
        I(X_2;Y) - I(\hY_1;Y_1|X_1,X_2,Y) \ge I(X_1;\hY_1|X_2,Y),
    \]
   hence both rate expressions, joint typicality and standard EAF, are identical.
   Now, consider a mapping $p(\hy_1|x_2,y_1)$ such that $I(X_1;\hY_1|X_2,Y) > 0$ and
   \begin{equation}
   \label{eqn:cond_joint}
            I(\hY_1;Y_1|X_1,X_2,Y) \le I(X_2;Y) \le
            I(\hY_1;Y_1|X_2,Y).
   \end{equation}
   In this situation, {\em assuming the same $p(\hy_1|x_2,y_1)$},
   joint decoding may provide a positive rate increase over the point-to-point rate $I(X_1;Y|X_2)$, while \cite[theorem 6]{CoverG:79} does not.
   The rate joint-decoding provides in this region is
   \[
        R  \le  I(X_1;Y|X_2) + I(X_2;Y) - I(\hY_1;Y_1|X_1,X_2,Y).
   \]
   Now we can define a time-sharing variable $\hat{\hY}_1$, $q \in [0,1]$
   \[
        p(\hat{\hy}_1|\hy_1) = \left\{
            \begin{array}{cl}
                q &,\hat{\hy}_1 = \hy_1\\
                1-q & ,\hat{\hy}_1 = \mbox{const} \notin \mhY_1.
            \end{array}
        \right.
   \]
   Then under the chain of distribution $p(x_1)p(x_2)p(y,y_1|x_1,x_2)p(\hy_1|x_2,y_1)p(\hat{\hy}_1|\hy_1)$,
     the joint decoding rate can be shown to be:
   \begin{eqnarray*}
        R & \le & I(X_1;Y|X_2)  + \min(I(X_2;Y) - I(\hhY_1;Y_1|X_1,X_2,Y),  I(X_1;\hhY_1|X_2,Y))\\
        \mbox{subject to }I(X_2;Y) & \ge & I(\hhY_1;Y_1|X_1,X_2,Y).
   \end{eqnarray*}
   Now using the above assignment of $p(\hat{\hy}_1|\hy_1)$ this can also be written as:
    \begin{eqnarray*}
        R & \le & I(X_1;Y|X_2)  + \min(I(X_2;Y) - qI(\hY_1;Y_1|X_1,X_2,Y),  qI(X_1;\hY_1|X_2,Y))\\
        \mbox{subject to }I(X_2;Y) & \ge & qI(\hY_1;Y_1|X_1,X_2,Y).
   \end{eqnarray*}
   When $q=1$ we have the original joint-decoding rate, since (in our example)
   \[
        I(X_2;Y) - I(\hY_1;Y_1|X_1,X_2,Y) \le   I(X_1;\hY_1|X_2,Y).
   \] However, decreasing $q$ we get that
    $I(X_2;Y) - qI(\hY_1;Y_1|X_1,X_2,Y)$ increases, while\\
     $qI(X_1;\hY_1|X_2,Y)$
    decreases. Therefore, time-sharing $\hY_1$ we improve upon the rate of joint decoding
    obtained without time-sharing $\hY_1$ in the region where joint-decoding is supposed to be better than EAF.
    Now, we keep decreasing $q$ until
    \begin{equation}
    \label{eqn:1}
        I(X_2;Y) - qI(\hY_1;Y_1|X_1,X_2,Y) =   qI(X_1;\hY_1|X_2,Y),
    \end{equation}
    and we get back to the original expression
    of \cite[theorem 6]{CoverG:79} with a higher rate than joint-decoding.
    Note that $I(X_2;Y) \ge I(\hY_1;Y_1|X_1,X_2,Y)$ implies that also
    $I(X_2;Y) \ge qI(\hY_1;Y_1|X_1,X_2,Y)$. If the solution to the equality is $q \ge 1$ then this implies that
    $I(X_2;Y) \ge I(\hY_1;Y_1|X_2,Y)$ and we are at the original \cite[theorem 6]{CoverG:79} situation to begin with.

    In conclusion, all the rates that joint decoding allows, can also be obtained or exceeded
    by the original EAF scheme with an appropriate time sharing variable (i.e. an appropriate mapping).
        This argument is due to Shlomo Shamai and Gerhard Kramer.

    Note that optimality in \eqref{eqn:1} implies
    \[
        q_{opt} = \min\left\{1,\frac{I(X_2;Y)}{I(\hY_1;Y_1|X_1,X_2,Y)+ I(X_1;\hY_1|X_2,Y)}\right\}
            =\min\left\{1, \frac{I(X_2;Y)}{I(\hY_1;Y_1|X_2,Y)}\right\}
            ,
    \]
    hence $q_{opt}$ is the maximum $q$ that makes the mapping $p(\hy_1|x_2,y_1)$ feasible for \cite[theorem 6]{CoverG:79}.
    Therefore, the rate
    \[
        R \le I(X_1;Y|X_2) + \min\left\{1, \frac{I(X_2;Y)}{I(\hY_1;Y_1|X_2,Y)}\right\}I(X_1;\hY_1|X_2,Y),
    \]
    is always achievable.
    Finally, consider again the region where joint decoding is useful \eqref{eqn:cond_joint}
    \begin{eqnarray*}
        I(\hY_1;Y_1|X_1,X_2,Y) & \le I(X_2;Y) \le &   I(\hY_1;Y_1|X_2,Y)\\
    \Rightarrow   0 & \le I(X_2;Y) - I(\hY_1;Y_1|X_1,X_2,Y) \le &   I(\hY_1;Y_1|X_2,Y) - I(\hY_1;Y_1|X_1,X_2,Y)\\
   \Rightarrow   0 & \le I(X_2;Y) - I(\hY_1;Y_1|X_1,X_2,Y) \le &    I(X_1;\hY_1|X_2,Y)\\
        \Rightarrow   0 & \le \frac{I(X_2;Y) - I(\hY_1;Y_1|X_1,X_2,Y)}{I(X_1;\hY_1|X_2,Y)} \le &   1.
    \end{eqnarray*}
    So, applying time-sharing on $\hY_1$ with:
    \[
        q = \frac{I(X_2;Y)-I(\hY_1;Y_1|X_1,X_2,Y)}{I(X_1;\hY_1|X_2,Y)},
    \]
    to \cite[theorem 6]{CoverG:79} yields:
    \[
        I(X_1;Y|X_2) + q I(X_1;\hY_1|X_2,Y) = I(X_1;Y|X_2) + I(X_2;Y)-I(\hY_1;Y_1|X_1,X_2,Y).
    \]
    Hence the joint-decoding rate can be obtained by time sharing on the \cite[theorem 6]{CoverG:79} expression.

In conclusion, since joint-decoding can be represented as a special case of time sharing
\cite[theroem 6]{CoverG:79}, we decided that it
does not merit a separate publication and we incorporated this into our work on the application of time-sharing to
estimate-and-forward.


\end{document}